\documentclass[journal,transmag]{IEEEtran}

\usepackage[T1]{fontenc} 
\usepackage{standalone}
\usepackage{ninecolors}
\usepackage{subcaption}%
\usepackage{amsmath}
\usepackage{siunitx}
\usepackage{acro}
\DeclareAcronym{fe}{short={FE}, long={finite element}}
\usepackage{siunitx}
\usepackage{tikz}
\usepackage{pgfplots}
\pgfplotsset{compat=1.18}
\pgfplotsset{table/search path={./tikz/data,./data}}
\usepackage{tikz-3dplot}
\usetikzlibrary{arrows.meta, calc, math}
\usetikzlibrary{fpu}
\usetikzlibrary{pgfplots.colormaps,patterns, patterns.meta}
\usepgfplotslibrary{colorbrewer, patchplots}
\pgfkeys{/pgf/fpu/output format=sci}
\usepackage{pgffor}
\pgfplotscreateplotcyclelist{mylist}{black,azure4,teal7}
\usepackage{tudacolors}
\pgfplotscreateplotcyclelist{myColorCycleList}{black,azure4,teal7,yellow8}
\usepackage{tabularray}
\usepackage[europeaninductors]{circuitikz}
\newcommand{\vect}[1]{\ensuremath{\vec{#1}}}
\newcommand{\Av}{\ensuremath{\vect{A}}}
\newcommand{\Jv}{\ensuremath{\vect{J}}}
\DeclareMathOperator{\gradword}{grad}

\DeclareMathOperator{\curlword}{curl}

\newcommand{\Grad}{\gradword}

\newcommand{\Curl}{\curlword}
\newcommand{\diff}{\ensuremath{\mathrm{d}}}
\newcommand{\trans}{^{\mkern-1.5mu\mathsf{T}}} %

\newcommand{\ds}{\diff s}

\newcommand{\dV}{\diff V}

\newcommand{\domain}{\Omega}
\newcommand{\fwdomain}{{\domain_{\mathrm{fw}}}}

\newcommand{\locala}{\alpha}
\newcommand{\localb}{\beta}
\newcommand{\localc}{\gamma}

\newcommand{\elocalc}{\vec{e}_{\localc}}

\newcommand{\Llocala}{L_{\locala}}

\newcommand{\widthconductor}[1][]{b_{\mathrm{c}#1}}

\newcommand{\foilwidth}{b}
\newcommand{\fillfactor}{\lambda}
\newcommand{\turns}{N}

\newcommand{\current}{I}
\newcommand{\voltage}{V}

\newcommand{\vdf}{\vec{\zeta}}
\newcommand{\volfun}{\Phi}

\newcommand{\mass}{\mathbf{M}}
\newcommand{\stiffness}{\mathbf{K}}
\newcommand{\coupling}{\mathbf{X}}
\newcommand{\conductance}{\mathbf{G}}
\newcommand{\cvec}{\mathbf{c}}
\newcommand{\qvec}{\mathbf{q}}

\newcommand{\masssigma}{\mass_{\sigma}}
\newcommand{\stiffnessnu}{\stiffness_{\nu}}
\newcommand{\couplingsigma}{\coupling_{\sigma}}
\newcommand{\conductancesigma}{\conductance_{\sigma}}

\newcommand{\w}{\vec{w}}

\newcommand{\sbf}{g}  %

\newcommand{\numdofAv}{N_{\mathrm{a}}}
\newcommand{\numdofvolfun}{N_{\mathrm{u}}}

\newcommand{\dS}{\diff S}

\newcommand{\adof}{\mathbf{a}}
\newcommand{\udof}{\mathbf{u}}

\newcommand{\Jsv}{\Jv_{\mathrm{s}}}

\newcommand{\dist}[1]{d_{\mathrm{}}}

\newcommand{\rv}{\vec{r}}

\usepackage[style=ieee,maxnames=1, minnames=1,nohashothers=true,isbn=false,doi=false,url=false,sorting=none]{biblatex}
\standaloneconfig{mode=image}

\usepackage{hyperref}
\hypersetup{hidelinks}

\begin{document}
	\title{Homogenization of Foil Windings with Globally Supported Polynomial Shape Functions}
	\author{\IEEEauthorblockN{Jonas~Bundschuh\IEEEauthorrefmark{1,2}, Yvonne Späck-Leigsnering\IEEEauthorrefmark{1,2}, and Herbert De Gersem\IEEEauthorrefmark{1,2}}
		\IEEEauthorblockA{\IEEEauthorrefmark{1}Institute for Accelerator Science and Electromagnetic Fields (TEMF), TU Darmstadt, 64289 Darmstadt, Germany}
		\IEEEauthorblockA{\IEEEauthorrefmark{2}Graduate School of Excellence Computational Engineering, 64293 Darmstadt, Germany}%
	}

	\IEEEtitleabstractindextext{%
		\begin{abstract}%
			In conventional finite element simulations, foil windings with thin foils and with a large number of turns require many mesh elements.
			This renders models quickly computationally infeasible.
			This paper uses a homogenized foil winding model and approximates the voltage distribution in the foil winding domain by globally supported polynomials. 
			This way, the small-scale structure in the foil winding domain does not have to be resolved by the finite element mesh. 
			The method is validated successfully for a stand-alone foil winding example and for a pot inductor example. Moreover, a transformer equipped with a foil winding at its primary side is simulated using a field-circuit coupled model.
		\end{abstract}
		}

	\maketitle

	\section{Introduction}
	\IEEEPARstart{F}{oil} windings are constructed by winding a thin, insulated metal foil around a support. Foil windings are preferred over wire windings because of their higher fill-factor, better thermal properties and lower costs \cite{Barrios_2015aa}.
	A foil winding may consist of hundreds of turns of a thin, conducting foil and even thinner insulation layers.
	
	The periodic structure of conducting and insulating materials in the direction perpendicular to the foils and an invariant geometry in the other directions cause a specific eddy current effect. 
	There exist analytical or semi-analytical methods to compute the eddy currents inside foil windings \cite{Kazimierczuk_2010aa}, \cite{Leuenberger_2015aa}.
	Nevertheless, the numerical field simulation of foil winding applications is indispensable because analytical models for arbitrary configurations do not exist.
	
	In a standard \ac{fe} procedure, the mesh resolves each foil and each insulation layer separately. 
	Due to the small dimensions of the single foils, this quickly leads to extremely large meshes and, thus, to prohibitively long simulation times \cite{Gyselinck2005a}.
	As a remedy, homogenization techniques have been developed \cite{Bossavit1994a}. 
	They model the electromagnetic phenomena on the basis of a comparatively coarse mesh.
	
	This paper formulates a homogenization approach for foil windings based on global polynomials to approximate the voltage variation in the direction perpendicular to the foils. 
	Within this approach, the foil windings are replaced by a homogenized material and an additional equation is expressed on the foil winding domain to model the individual windings to carry the same current \cite{Valdivieso_2021aa}.
	The homogenization technique is validated for two academic examples and illustrated for a pot transformer.
	
	\section{Formulation}
\begin{figure}
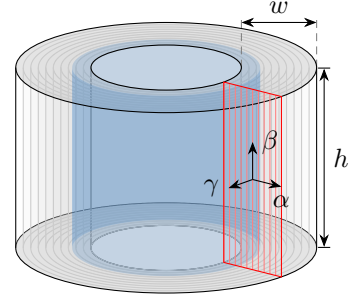

	\centering
	\includestandalone{schematic}
	\caption{Schematic representation of an exemplary foil winding domain $\fwdomain$. The coordinates $\locala$, $\localb$ and $\localc$ are perpendicular to the foils, in the direction of the tips and in winding direction, respectively. The constant cross section is highlighted in red and the surface $\Gamma(\locala)$ is illustrated for a fixed $\locala$ in blue.}
	\label{fig:schematic}
\end{figure}
The problem is formulated in the magnetoquasistatic re\-gime, using the magnetic vector potential $\Av(\rv,t)$ and the electric scalar potential $\phi(\rv,t)$. 
In \cite{De-Gersem_2001aa}, the foil winding was homogenized into a \emph{foil conductor model} for 2D Cartesian models. The method was extended to 3D in \cite{Dular_2002aa} and to the 2D axisymmetric case in \cite{Valdivieso_2021aa}. This paper follows the general derivation provided in \cite{Paakkunainen_2023}. The homogenized model is based on a \emph{voltage function} $\volfun(\locala,t)$ which only depends on the spatial coordinate perpendicular to the foils, here denoted by $\locala$, and on a \emph{distribution function} $\vdf(\rv)$ which is oriented along the winding direction $\elocalc$ and integrates up to $1$ when integrated along the circumference of the foil winding (see Fig.~\ref{fig:schematic}). They are related via $-\Grad\phi=\volfun\vdf$. The homogenized system reads
\begin{subequations}\label{eq:system_equations}	
	\begin{align}
		\Curl\left(\nu\Curl\Av\right) + \sigma\partial_t\Av - \sigma\volfun\vdf &= \Jsv\,, && \text{in }\domain,\label{eq:system_equations_a}\\
		\int_{\Gamma(\locala)} \sigma\left(-\partial_t\Av + \volfun\vdf\right)\cdot\vdf \,\dS &= \frac{\current}{\foilwidth}\,, && \text{in }\Llocala,\label{eq:system_equations_b}
	\end{align}
\end{subequations}
where $\nu$ and $\sigma$ are the reluctivity and the conductivity, and where $\domain$ is the entire computational domain. In the foil winding domain $\fwdomain\subseteq\domain$, $\nu$ and $\sigma$ are homogenized using mixing rules \cite{Paakkunainen_2023}. $\Llocala$ denotes the domain spanned by the perpendicular coordinate $\locala$ and $\Gamma(\locala)$ denotes the surface for constant $\locala$ (see Fig.~\ref{fig:schematic}) \cite{Paakkunainen_2023}. $\Jsv$ models excitations in wire windings and solid conductors. The formulation is completed by suitable boundary conditions on $\partial\domain$ and initial values at a time point $t_0$.

	\section{Spatial Discretization}
	The homogenized system of equations \eqref{eq:system_equations} is discretized in space using the Galerkin approach. The magnetic vector potential $\Av(\rv,t)$ is approximated with standard \ac{fe} edge shape functions $\w_j(\rv)$, whereas the voltage function $\volfun(\locala,t)$ is approximated with scalar functions $\sbf_j(\locala)$, i.e.,
	\begin{subequations}
		\begin{align}\label{eq:field_with_basis}
			\Av(\rv,t) &= \sum_{j=1}^{\numdofAv}a_j(t)\w_j(\rv)\,, \\
			\volfun(\locala,t) &= \sum_{j=1}^{\numdofvolfun}u_j(t)\sbf_j(\locala)\,.
		\end{align}
	\end{subequations}
	The functions $\sbf_j(\locala)$ are nonzero only in the foil winding domain, depend on the coordinate $\locala$ perpendicular to the foils and are constant in $\Gamma(\locala)$.
	
	The Ritz-Galerkin method leads to the set of differential-algebraic equations
	\begin{subequations}
		\begin{align}
			\stiffnessnu \adof + \masssigma\partial_t\adof - \couplingsigma\udof &= \qvec\,,\\
			-\couplingsigma\trans\partial_t\adof + \conductancesigma\udof - I\cvec &= 0\,,\\
			\cvec\trans\udof &= \voltage,
		\end{align}
	\end{subequations}
	with the voltage drop $\voltage$ over the foil winding and the current $\current$ through the foil winding. The matrices are defined by
	\begin{subequations}
		\begin{align}
			\left[\stiffnessnu\right]_{i,j} &= \int_{\domain} \nu \Curl\w_j \cdot \Curl\w_i\,\dV\,,\\
			\left[\masssigma\right]_{i,j} &= \int_{\domain} \sigma \w_j \cdot \w_i \,\dV\,,\\
			\label{eq:mixed}\left[\couplingsigma\right]_{i,j} &= \int_{\domain} \sigma \sbf_j \vdf\cdot\w_i\,\dV\,,\\
			\left[\conductancesigma\right]_{i,j} &= \int_{\domain} \sigma \vdf\cdot\vdf \sbf_j \sbf_i \,\dV\,,\\
			\left[\qvec\right]_{i} &= \int_{\Omega} \Jsv\cdot\w_i\,\dV\,,\\
			\left[\cvec\right]_{i} &= \frac{1}{\foilwidth} \int_{\Llocala} \sbf_i \,\ds\,.    
		\end{align}
	\end{subequations}
	If the field model is coupled to a circuit model, the matrix system can be augmented with a matrix system obtained by modified nodal analysis of the circuit \cite{Schops_2013aa}.

	\section{Discretization of the voltage function}
	Several choices of shape functions $\sbf_j(\locala)$ for the voltage function are possible. 
	
	\subsection{Hat functions}
	\begin{figure}
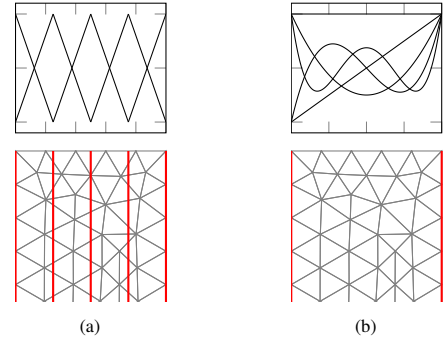

		\centering
		\begin{subfigure}{.4\columnwidth}
			\centering
			\includestandalone{mesh_intersection_hat}
			\caption{}
			\label{fig:mesh_intersection_hat}
		\end{subfigure}
		\begin{subfigure}{.4\columnwidth}
			\centering
			\includestandalone{mesh_intersection_poly}
			\caption{}
			\label{fig:mesh_intersection_poly}
		\end{subfigure}
		\caption{Basis functions $\sbf_j$ over the perpendicular direction on the foil winding domain (top). The 1D mesh for the $\sbf_j$ is in red and a section of the triangular \ac{fe} mesh is in gray (bottom). For hat functions \subref{fig:mesh_intersection_hat} the meshes intersect each other, while for the polynomials \subref{fig:mesh_intersection_poly} the foil winding region is considered as a single element.}
		\label{fig:mesh_intersection}
	\end{figure}
	
	In \cite{De-Gersem_2001aa}, hat functions are used for $\sbf_j$ (see Fig.~\ref{fig:mesh_intersection_hat}). Thanks to their compact support, $\conductancesigma$ and $\couplingsigma$ become sparse matrices. However, the edge functions $\w_j(\rv)$ and the hat functions are defined on two different, intersecting meshes. This hampers the calculation of the mixed integrals \eqref{eq:mixed} considerably. An exact evaluation of the integrals necessitates the construction of the intersecting mesh as is done in \cite{De-Gersem_2001aa}, which is tedious. As an alternative, the mixed integrals can be evaluated by Gauss quadrature on the mesh of $\w_j$. This sacrifices the beneficial properties of Gauss quadrature because the hat functions are not infinitely differentiable within the simplices of that mesh. A high integration order may increase the accuracy but does not fully restore the convergence order of the Gauss quadrature schemes. This causes hat functions to be impractical for real problems involving foil windings.
	
	\subsection{Globally supported polynomials}
	In this paper, $\sbf_j$ are chosen to be polynomials with a global support in $\Llocala$. Then, $\couplingsigma$ contains dense blocks according to the foil winding domains. For two reasons, the computation turns out to be simpler compared to the one for the hat functions.
	First, the polynomials are infinitely differentiable. Second, there is no need for mesh intersection (see Fig.~\ref{fig:mesh_intersection_poly}). Since the $j$-th column of $\couplingsigma$ has the same form as the vector $\qvec$ with $\sigma\sbf_j\vdf$ as artificial current density, the routine for the computation of $\qvec$ can also be used for the computation of $\couplingsigma$. In the following sections, Legendre polynomials are used for $\sbf_j$.

	\section{Validation}
	We validate the homogenization with two examples: a stand-alone foil winding in Cartesian coordinates with flux wall boundary conditions and a foil winding as part of an axisymmetric pot inductor with an air gap. Both examples are chosen such that an analytical solution can be computed \cite{Bundschuh_2022ac}.
	The homogenization ansatz is implemented in the Python-based FE simulation framework \emph{Pyrit} \cite{Bundschuh_2022ab}.
	
	\subsection{First validation: Stand-alone foil winding}
	\begingroup
	\pgfkeys{/pgf/fpu=true}
	\tikzmath{
		integer \conductivity, \conductivityMSm, \frequency, \xplwidthmm, \xplheightmm, \xplturns, \frequencykHz;
		\xplwidthmm = 2;  %
		\xplheightmm = 4;  %
		let \xplfillfactor = 0.9;
		let \xplturns = 100;
		\conductivity = 5.7e7;
		\conductivityMSm = 57;
		let \frequencykHz = 50;
		\frequency = \frequencykHz * 1000;
		\angfrequency = 2*pi*\frequency;
		\permeability = 1 * 4 * pi * 1e-7;
		\xplskindepth = sqrt(2/(\angfrequency*\permeability*\conductivity));
		\xplskindepthmm = 1000*\xplskindepth;  %
		\xplfoilwidthmm = \xplwidthmm/\xplturns;
		\xplwidthconductormm = \xplfillfactor * \xplwidthmm/\xplturns;
		\xplskindepthtowidthconductor = \xplskindepthmm/\xplwidthconductormm;  %
		\xplskindepthtoheight = \xplskindepthmm/(\xplheightmm);  %
	}
	\pgfkeys{/pgf/fpu=false}
	We consider a foil winding of width $w=\qty{\xplwidthmm}{\milli\meter}$ and height $h=\qty{\xplheightmm}{\milli\meter}$. The model is in Cartesian coordinates and has a length of \qty{500}{\milli\meter}. Consequently, we can reduce the simulation to the two-dimensional cross section of the model. 
	Furthermore, the foil winding has $\turns=\num[round-precision=1]{\xplturns}$ turns and a fill factor of $\fillfactor=\num[round-mode=none]{\xplfillfactor}$. 
	This leads to a foil width $\foilwidth=\qty{\xplfoilwidthmm}{\milli\meter}$ and conductor width $\widthconductor=\qty{\xplwidthconductormm}{\milli\metre}$. 
	The foil winding is made out of a copper foil with conductivity $\sigma=\qty{\conductivityMSm}{\mega\siemens\per\meter}$. The simulations are performed at a frequency of $f=\qty{\frequencykHz}{\kilo\hertz}$, which leads to a skin depth of $\delta\approx\qty{\xplskindepthmm}{\milli\meter}$. 
	The ratios $\frac{\delta}{\widthconductor}\approx\num{\xplskindepthtowidthconductor}$ and $\frac{\delta}{h}\approx\num{\xplskindepthtoheight}$ show that eddy currents can be neglected perpendicular to the foils but are relevant in the direction of the tips.
	
	Figure \ref{fig:conv_rectangle} shows the relative error in the magnetic energy $W$ with respect to the analytical magnetic energy $W_{\mathrm{ana}}$ for increasingly finer \ac{fe} meshes and for different numbers $\numdofvolfun$ of hat functions and polynomials. For both choices, the energy converges against the analytical solution. 
	However, the relative error stagnates at a certain value with hat functions. 
	To reach the same accuracy, less polynomials are needed than hat functions. Here, even three polynomials have a lower error than six hat functions. The cases with one polynomial and with one hat function are identical by construction. 
	
	\begin{figure}
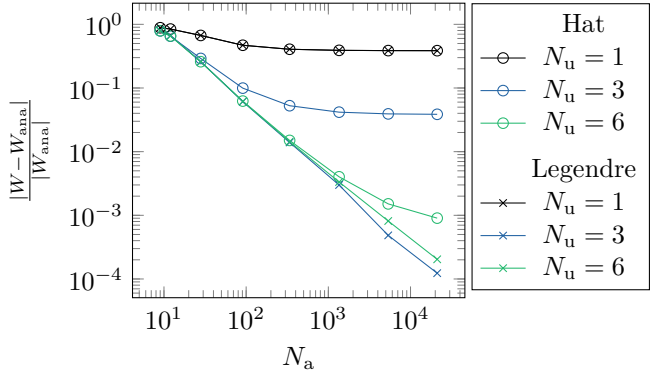

		\centering
		\includestandalone{convergence_plot}
		\caption{Relative error in the magnetic energy $W$ with respect to the analytical magnetic energy $W_{\mathrm{ana}}$ over the number $\numdofAv$ of degrees of freedom for the magnetic vector potential for the stand-alone foil winding. There are plots for different numbers $\numdofvolfun$ of scalar basis functions, for hat functions (circles) and polynomials (crosses).}
		\label{fig:conv_rectangle}
	\end{figure}
	
	\endgroup
	
	\subsection{Second validation: Pot inductor}
	\begingroup
	\pgfkeys{/pgf/fpu=true}
	\tikzmath{
		integer \conductivity, \conductivityMSm, \frequency, \xplwidthmm, \xplheightmm, \xplturns, \frequencykHz;
		\xplwidthmm = 30;  %
		\xplheightmm = 10;  %
		let \xplfillfactor = 1;
		let \xplturns = 200;
		\conductivity = 5.7e7;
		\conductivityMSm = 57;
		let \frequencykHz = 10;
		\frequency = \frequencykHz * 1000;
		\angfrequency = 2*pi*\frequency;
		\permeability = 1 * 4 * pi * 1e-7;
		\xplskindepth = sqrt(2/(\angfrequency*\permeability*\conductivity));
		\xplskindepthmm = 1000*\xplskindepth;  %
		\xplfoilwidthmm = \xplwidthmm/\xplturns;
		\xplwidthconductormm = \xplfillfactor * \xplwidthmm/\xplturns;  %
		\xplskindepthtowidthconductor = \xplskindepthmm/\xplwidthconductormm;  %
		\xplskindepthtoheight = \xplskindepthmm/(\xplheightmm);  %
	}
	\pgfkeys{/pgf/fpu=false}
	
	For the second validation example, we simulate a pot inductor. Figure~\ref{fig:geometry_pot_inductor} shows a cut of the axially symmetric domain for positive radii. It consists of a yoke in dark gray, the foil winding inside of the yoke in light gray and an air gap in white. The dimensions can be taken from the figure. Note that the windings are extended to the yoke. This is not realistic but allows to construct an analytical reference solution \cite{Bundschuh_2022ac}. For this example, we use a disk type foil winding, i.e. where the perpendicular direction coincides with the axial direction, with $\turns=\num[round-precision=1]{\xplturns}$ turns and at a frequency of $f=\qty{\frequencykHz}{\kilo\hertz}$. 
	
	In the analytical solution and in the simulations for the convergence plot, we assume for the permeability of the yoke $\mu\to\infty$. This allows to only consider the regions inside the yoke, i.e. the air gap and the foil winding, since the tangential component of the magnetic field then has to vanish at the surface of the yoke. 
	Figure~\ref{fig:conv_pot_inductor} shows the convergence of the magnetic energy $W$ of the pot inductor for hat functions and polynomials. Like in the previous example, the polynomials show a better convergence.
	\begin{figure}
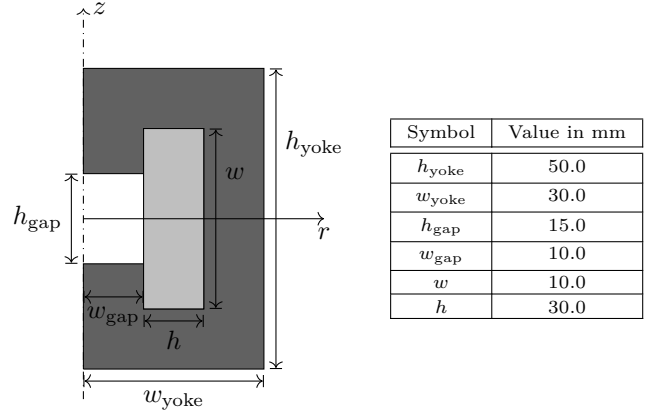

		\centering
		\includestandalone{geometry_pot_inductor}
		\caption{Geometry and dimensions of the pot inductor, used for the second validation example.}
		\label{fig:geometry_pot_inductor}
	\end{figure}
	\begin{figure}
		\centering
		\includestandalone{convergence_pot_inductor}
		\caption{Relative error in the magnetic energy $W$ with respect to the analytical magnetic energy $W_{\mathrm{ana}}$ over the number $\numdofAv$ of degrees of freedom for the magnetic vector potential for the pot inductor. There are plots for different numbers $\numdofvolfun$ of scalar basis functions, for hat functions (circles) and polynomials (crosses).}
		\label{fig:conv_pot_inductor}
	\end{figure}
	\endgroup

	\section{Example: Pot transformer}
	
	\begin{figure}
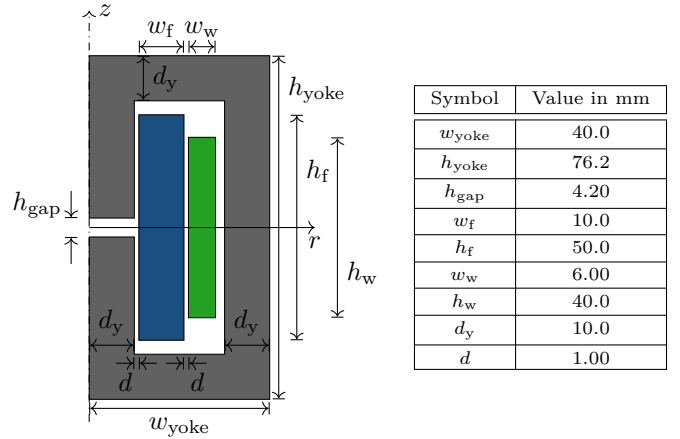

		\centering
		\includestandalone{pot_transformer_model}
		\caption{Geometry and dimensions of the pot transformer. The yoke is in dark gray and the air is in white. The inner winding (in blue) is a foil winding and the outer winding (in green) is a wire winding.}
		\label{fig:trafo_geometry}
	\end{figure}
	
	\begin{figure}
		\centering
		\includestandalone{pot_transformer_circuit}
		\caption{Surrounding circuit including the pot transformer. The values of the passive electrical components are $R=\qty{1}{\ohm}$, $R_{\mathrm{L}}=\qty{10}{\ohm}$ and $C=\qty{0.1}{\milli\farad}$.}
		\label{fig:trafo_circuit}
	\end{figure}
	
	The homogenization has been validated for two academic configurations. In the following, we show the homogenization for a more realistic example. We simulate a pot transformer including a surrounding circuit, as shown in Figs.~\ref{fig:trafo_geometry} and \ref{fig:trafo_circuit}.
	A tube type foil winding, i.e. where the perpendicular direction coincides with the radial direction, with $\turns=\num{100}$ turns and a fill factor of $\fillfactor_1=\num[round-precision=1]{0.8}$ is used for the primary side (blue region in Fig.~\ref{fig:trafo_geometry}). For the secondary side, a wire winding with $\turns=\num{500}$ turns and a fill factor of $\fillfactor_2=\num[round-precision=1]{0.8}$ is used (green region in Fig.~\ref{fig:trafo_geometry}). The dimensions of the pot transformer can be found in Fig.~\ref{fig:trafo_geometry}. 
	
	The simulation consists of a field problem, including the homogenized foil winding model and the stranded conductor model for the wire winding, and a circuit problem, including a voltage source, a capacitor and resistors. 
	It uses a field-circuit coupling, such that both problems are solved simultaneously. The excitation voltage is a square wave
	\begin{equation}
		V_{\mathrm{s}}(t) = 2\left\lfloor \frac{t}{T} - 0.25 \right\rfloor - \left\lfloor 2\left(\frac{t}{T} - 0.25\right) \right\rfloor + 1\,,
	\end{equation}
	with a period $T=\qty{20}{\milli\second}$.
	The simulation is carried out in time domain, with a backward Euler method used for time discretization. 
	
\begin{figure}
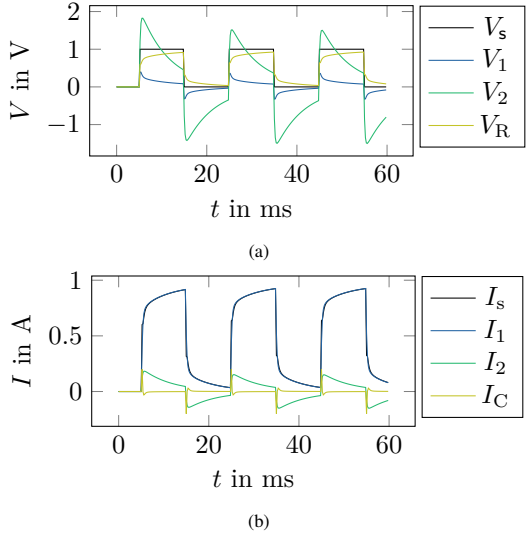

		\centering
		\begin{subfigure}{.8\columnwidth}
			\centering
			\includestandalone{plots_voltage}
			\caption{}
			\label{fig:voltages}
		\end{subfigure}
		\begin{subfigure}{.8\columnwidth}
			\centering
			\includestandalone{plots_current}
			\caption{}
			\label{fig:currents}
		\end{subfigure}
		\caption{Voltages \subref{fig:voltages} and currents \subref{fig:currents} over the time of three periods.}
		\label{fig:voltages_and_currents}
	\end{figure}

	Figures~\ref{fig:voltages} and \ref{fig:currents} show the voltages and currents in the circuit over time, respectively. The voltages and currents at the transformer, i.e. $V_1$, $V_2$, $I_1$ and $I_2$, are extracted directly from the degrees of freedom. The remaining ones are computed in a post-processing step with the known circuit relations. 
	
	\section{Conclusion}
	A homogenized foil winding model has been equipped with a better discretization of the voltage function based on polynomials with global support. As shown by two validation examples, the convergence of the polynomials outperforms the discretization based on hat functions. Moreover, the new approach considerably simplifies the calculation of the mixed integrals combining the edge functions for the magnetic vector potential with the shape functions for the voltage function. The method is illustrated for a transient field-circuit coupled model of a transformer with a foil winding at its primary side.

	\section*{Acknowledgment}
	The work is supported by the German Science Foundation (DFG project 436819664), the Graduate School CE within the Centre for Computational Engineering at the Technische Universität Darmstadt, and the Athene Young Investigator Fellowship of the Technische Universität Darmstadt.

	\printbibliography

\end{document}